\documentclass[pre,superscriptaddress,twocolumn]{revtex4-1}
\usepackage{amsfonts,amsmath,amssymb,graphicx,hyperref, multirow, longtable,siunitx, xr, color, CJK}

\newcommand{\tabincell}[2]{\begin{tabular}{@{}#1@{}}#2\end{tabular}}

\begin{document}
\begin{CJK*}{UTF8}{}
\title{Knowledge Evolution in Physics Research: An Analysis of Bibliographic Coupling Networks}
\author{Wenyuan Liu (\CJKfamily{gbsn}{刘文源})}
\affiliation{
   School of Physical and Mathematical Sciences,\\
   Nanyang Technological University,\\
   Singapore 637371, Singapore
   }
\author{Andrea Nanetti}
\affiliation{
   School of Art, Design and Media,\\
   Nanyang Technological University,\\
   Singapore 637371, Singapore
   }
\author{Siew Ann Cheong}
\email{cheongsa@ntu.edu.sg}
\affiliation{
   School of Physical and Mathematical Sciences,\\
   Nanyang Technological University,\\
   Singapore 637371, Singapore
   }

\date{\today}
\begin{abstract}
Even as we advance the frontiers of physics knowledge, our understanding of how this knowledge evolves remains at the descriptive levels of Popper and Kuhn. Using the APS publications data sets, we ask in this letter how new knowledge is built upon old knowledge. We do so by constructing year-to-year bibliographic coupling networks, and identify in them validated communities that represent different research fields. We then visualize their evolutionary relationships in the form of alluvial diagrams, and show how they remain intact through APS journal splits. Quantitatively, we see that most fields undergo weak Popperian mixing, and it is rare for a field to remain isolated/undergo strong mixing. The sizes of fields obey a simple linear growth with recombination. We can also reliably predict the merging between two fields, but not for the considerably more complex splitting. Finally, we report a case study of two fields that underwent repeated merging and splitting around 1995, and how these Kuhnian events are correlated with breakthroughs on BEC, quantum teleportation, and slow light. This impact showed up quantitatively in the citations of the BEC field as a larger proportion of references from during and shortly after these events.
\end{abstract}

\maketitle
\end{CJK*}

According to Karl Popper, science progresses through repeated hypothesis testing~\cite{Popper1999}.
Hypotheses contrary to empirical evidence must be rejected, while those consistent with data survive to be tested another day.
In this picture of the scientific enterprise, our knowledge of the world around us is always tentative, but becomes more complete over time.
On the other hand, Thomas Kuhn believes that the accepted knowledge of a given time is the result of consensus amongst scientists, based on evidences consistent with their theories~\cite{Kuhn1970}.
However, when too many conflicting evidences are found, a new consensus can form around new theories in what he called a ``paradigm shift''.
Kuhn gives special relativity and quantum theory as examples of paradigm shifts.
Looking back, we realize these two theories have enormous impacts on how we understand the world today.
But could there be paradigm shifts of various scales that have also contributed to reshaping our knowledge of physics?

Many historians of science have noted the strongly reductionistic flavor of scientific research in the last couple of centuries\cite{wootton2015invention}.
Starting as natural philosophy, the body of scientific knowledge became separated disciplines of astronomy, biology, chemistry and physics.
Within physics itself, we also observe the emergence of high energy physics, condensed matter physics, biophysics, and photonics.
These are the results of the splitting of science into more specialized fields.
We also observe in parallel the merging of fields, such as the merging of astronomy and physics to give astrophysics, biology and chemistry to give biochemistry, and others ``that arose by division and recombination of specialties already matured"~\cite{Kuhn1970}.
These developments have been discussed extensively by philosophers and historians of science, but unlike our quantitative understanding of physics, our appreciation for the processes through which we acquired our knowledge of physics remains at a highly descriptive level.
Some progress has been made in addressing this problem~\cite{Bollen2009, Kuhn2014, Jia2017}.
In particular, the following three papers provide the inspiration for our study.
Chen and Redner suggested that long-range connections can form between disparate fields because of the development of ``a widely applicable theoretical technique, or cross fertilization between theory and experiment''~\cite{Chen2010}.
Visualizing the cross citations between neuroscience journals, Rosvall and Bergstrom traced the growth and maturation of neuroscience as a discipline~\cite{Rosvall2010}.
Using embryology as a specific example, Chavalarias and Cointet created a phylomemetic network visualization for the evolution of science~\cite{Chavalarias2013}.

\begin{figure}
  \centering
  \includegraphics[width=\columnwidth]{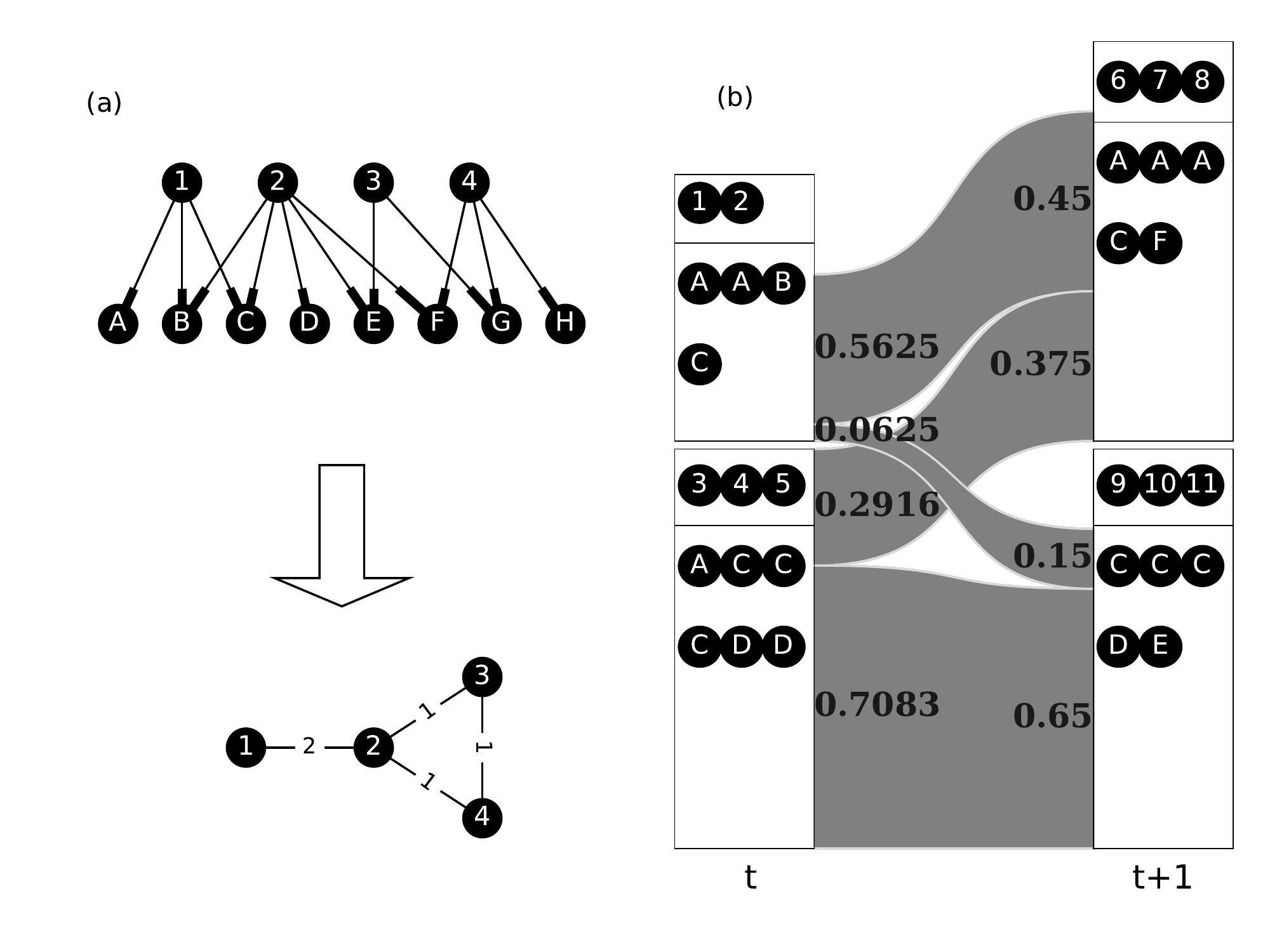}
  \caption{(a) Building a BCN (lower) from a citation network (upper): circles with numbers are papers under consideration, circles with letters are their references, and numbers on edges are weights. (b) Topical clusters in year $t$ (left) and in year $t+1$ (right), and their forward (left) and backward (right) intimacy indices, shown as flows. }\label{fig:example}
\end{figure}

While these previous studies point to the evolution of scientific knowledge, they do not identify the entity that is recognizably `knowledge', or they do not study the interactions between such objects.
To clarify what constitutes knowledge, we start with the bibliographic coupling network (BCN)~\cite{Kessler1963}, proposed by Kessler and used extensively in computer science~\cite{Yan2012, Huang2003}.
In a BCN, nodes represent papers, and if two papers share $w$ common references, we draw an edge with weight $w$ between them (see \autoref{fig:example}(a)).
The BCN is suitable for our purpose for two reasons: (i) the BCN for a given year consists only of papers published that year and does not change after more papers are published later, so features in the BCN represent the state of knowledge in that year;
and (ii) the appropriate collective unit of knowledge is a community in the BCN instead of a few key papers or a journal.
For the American Physical Society (APS) data set, consisting of about half a million publications between 1893 and 2013 \cite{APS_dataset}, we show in Supplementary Information \autoref{sec:nullmodel} that the BCN edge weights are far more heterogeneous than expected from an appropriate null model.

This heterogeneity can be explained by the presence of communities that we extracted using the modularity-optimized Louvain method~\cite{Blondel2008}.
Using the null model in \autoref{sec:nullmodel}, we show that these communities are statistically significant.
We also test in \autoref{sec:pacstest} how likely the most common PACS number in a community of $n$ papers can appear with its observed frequency, within random collections of $n$ papers.
For most communities, this is highly unlikely, so we conclude that the groupings of papers extracted are meaningful.
We refer to these validated units of knowledge as \emph{topical clusters} (TCs).

To study how knowledge evolves, we investigate how TCs $\{\mathcal{C}^t\}$ in year $t$ become $\{\mathcal{C}^{t+1}\}$ in year $t+1$.
The papers published in different years are distinct, but they do overlap in their references.
Therefore we use this fact to define a \emph{forward intimacy index} $I^{f}_{mn}$ and a \emph{backward intimacy index} $I^{b}_{mn}$:
\begin{equation}\label{eq:intimate_index}
  \begin{aligned}
    I^f_{mn} &= \sum_{i} \frac{N\left(R_i, \mathcal{R}_n^{t+1}\right)}{N\left(R_i, \mathcal{R}^{t+1}\right)} N\left(R_i, \mathcal{R}_m^t\right) \mathbin{/} L\left(\mathcal{R}_m^t\right); \\
    I^b_{mn} &= \sum_{i} \frac{N\left(R_i, \mathcal{R}_m^t\right)}{N\left(R_i, \mathcal{R}^{t}\right)} N\left(R_i, \mathcal{R}_n^{t+1}\right) \mathbin{/} L\left(\mathcal{R}_n^{t+1}\right),
  \end{aligned}
\end{equation}
to quantify how close $C^t_m$ is to $C^{t+1}_n$.
Here the TCs at $t$ and $t+1$ are $\mathcal{C}^t = \left\{C_1^t, ...,C_m^t,..., C_u^t\right\}$ and $\mathcal{C}^{t+1} = \left\{C_1^{t+1}, ...,C_n^{t+1},..., C_v^{t+1}\right\}$,
and we denote the references cited by papers in $C^t_m$  and $C^{t+1}_n$ as $\mathcal{R}_m^t = \mathcal{R}(C_m^t) = \left[R^m_1, ..., R^m_p\right]$ and $\mathcal{R}_n^{t+1} = \mathcal{R}(C_n^{t+1}) = \left[R^n_1, ..., R^n_q\right]$; and $\mathcal{R}^t = \left\{\mathcal{R}_1^t, ...,\mathcal{R}_m^t,...\right\}$.
$N(element, list)$ is the number of times $element$ occurs in $list$, and $L(list)$ is the length of $list$.
In this definition, we assume each citation instance in $t$ will be uniformly distributed over all instances of the same citation in $t+1$, while each citation in $t+1$ receives equal contributions from all instances of the same citation in $t$.
In general, this index is asymmetric, i.e. $I^f_{mn} \neq I^b_{mn}$, because the references are not cited the same number of times in the two years, as illustrated in \autoref{fig:example}(b).

\begin{figure*}
  \centering
  \includegraphics[width=\textwidth]{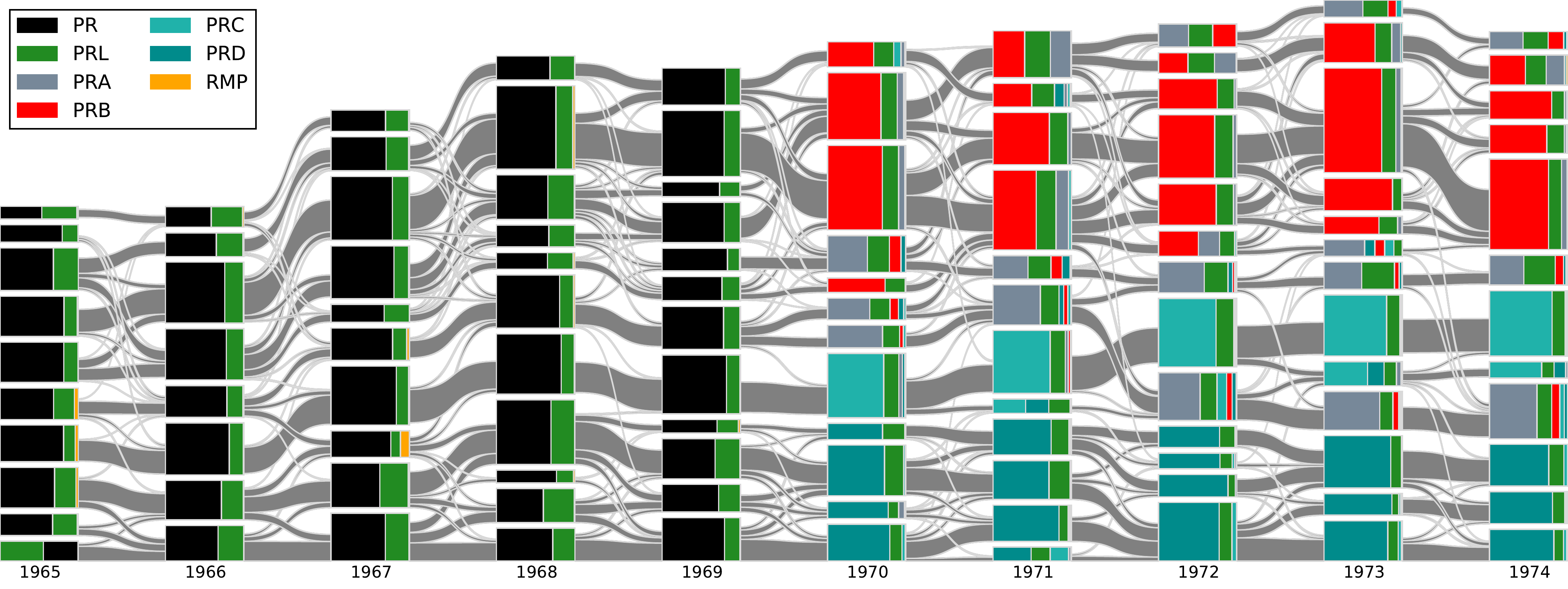}
  \caption{The alluvial diagram of APS papers from 1965 to 1974. Each block in a column represents a TC and the height of the block is proportional to the number of papers in the TC. Only communities comprising more than 100 papers are shown. TCs in successive years are connected by streams whose widths at the left and right ends are proportional to the forward and backward intimacy indices. The different colors in a TC represent the relative contributions from different journals.}\label{fig:alluvial_diagram}
\end{figure*}

We visualize the sequence of TCs and their intimacy indices,  the evolution of physics research they represent in the form of alluvial diagrams.
For example, in \autoref{fig:alluvial_diagram} we can clearly see the birth of PRA, PRB, PRC and PRD from PR in 1970.
Each journal consist of several TCs, which existed even in the PR era.
The editorial decision to split PR is consistent with the self-organized TCs even though it was done without classification analysis.
We also see the consistent birth of PRE from PRA in 1993 In \autoref{sec:prebirth}.

More importantly, from the alluvial diagram we can identity the key interactions between TCs that are correlated with important publications.
In \autoref{fig:BEC} of \autoref{sec:BEC} we showcase one such episode between 1991 and 2000.
At the beginning of the decade, we see two PRA-dominated TCs.
Based on the papers they contain, we can loosely associate one with quantum information (QI) and trapped atomic ions (BEC), and the other with quantum optics (QO).
In 1993, the QI + BEC TC cited many QO papers, and in 1994, the QO TC cited many QI + BEC papers.
Following this `cross-fertilization', the two TCs merged in 1995, the same year Cornell \emph{et al.} \cite{Anderson1995} and Ketterle \emph{et al.} \cite{Davis1995} published their seminal papers demonstrating BEC in dilute atomic gases.
In recognition of their works, Cornell, Wieman, and Ketterle were awarded the 2001 Nobel Prize in Physics.
The PRA-dominated TC split after 1996 to give one that is exclusively BEC, and another that is still a combination of QI + QO.
It was after Zeilinger demonstrated in 1997 experimental quantum teleportation \cite{Bouwmeester1997} that the QI + QO TC split into a QI TC and a QO TC.
After receiving more influence from other PRB-dominated TC, the QO cluster produced yet another breakthrough paper, in the form of ultraslow light in hot atomic gases\cite{Kash1999}.
Without the data visualization done here, few may suspect the existence of such connections between BEC, quantum teleportation and slow light.

From \autoref{fig:alluvial_diagram} we see a diversity of inflows and outflows from one TC to another: some TCs are derived almost exclusively from one source, others receive strong contributions from a small number of sources, or weak contributions from a large number of sources.
To quantify such diversity, we construct a forward mixing degree of community $C_m^t$ and backward mixing degree of $C_n^{t+1}$ analogous to the Gini-Simpson index~\cite{Jost2006}:
\begin{equation}\label{eq:mixingdegree}
  \begin{aligned}
    M_m^f & = 1 - \textstyle\sum_{n} \left(I_{mn}^f \mathbin{/} \textstyle\sum_{n'} I_{mn'}^f\right)^2,\\
    M_n^b & = 1 - \textstyle\sum_{m} \left(I_{mn}^b \mathbin{/} \textstyle\sum_{m'} I_{m'n}^b\right)^2.
  \end{aligned}
\end{equation}
A TC with low forward/backward mixing degree has effectively one child/parent, whereas a TC with high forward/backward mixing degree undergoes/results from strong splitting/merging.
As shown in \autoref{sec:metabolism}, neither are frequent. It is more common to find weak mixing between TCs, which we believe is due to most papers citing small numbers of papers outside their fields.

\begin{figure}
  \centering
  \includegraphics[width=\columnwidth]{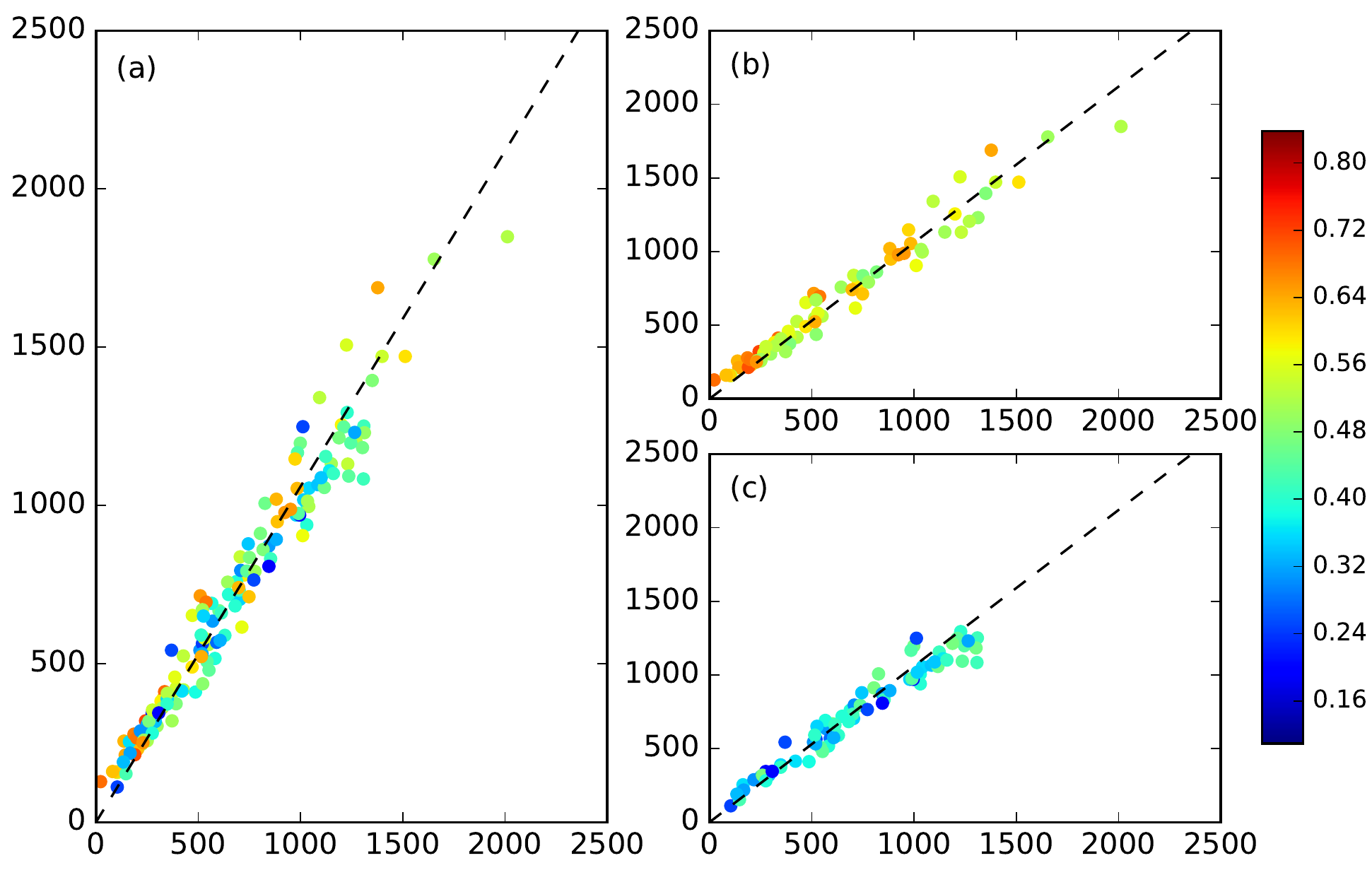}
  \caption{(a) Plot of observed (y-axis) against predicted (x-axis) sizes of recombined TCs, showing a linear growth with slope 1.06 (dashed line). This linear growth is the same for TCs with (b) high (red) or (c) low (blue) backward mixing degree.}\label{fig:sizerelation}
\end{figure}

At this point, let us recall the Popperian and Kuhnian pictures of the evolution of knowledge, where we expect incremental growth punctuated by abrupt paradigm shifts.
Certainly, at the aggregate level of PR series of premier physics journals, the number of articles published has grown over the years.
When we partition these articles into TCs, we naively expect that some clusters will grow/shrink because of growing/declining interest in their topics.
From the alluvial diagrams, we realize that the real picture is far more complex because of recombinations between TCs.
Therefore, instead of measuring the growth rates of pure TCs, we need to measure the growth of recombined TCs.
To do this, we assume that the contribution of $C^t_m$ to the size of $C^{t+1}_n$ is proportional to the size of $C^t_m$ and also the normalized forward intimacy index $I^f_{mn} \mathbin{/} \sum_{n} I_{mn}^f$, i.e.
\begin{equation}\label{eq:sizeprediction}
  L^\prime(C_n^{t+1}) = \textstyle\sum_{m} L(C_m^t) (I_{mn}^f \mathbin{/} \textstyle\sum_{n} I_{mn}^f).
\end{equation}
When we plot the predicted sizes $L^\prime(C_n^{t+1})$ against the observed size $L(C_n^{t+1})$ in \autoref{fig:sizerelation}, we find $(L'(C_n^{t+1}), L(C_n^{t+1}))$ scattered about about a straight line with slope with 1.06, which is the annual growth rate of the number of papers in APS journals.
This tells us that the growth of recombined TCs is also Popperian.

\begin{figure}
  \centering
  \includegraphics[width=\columnwidth]{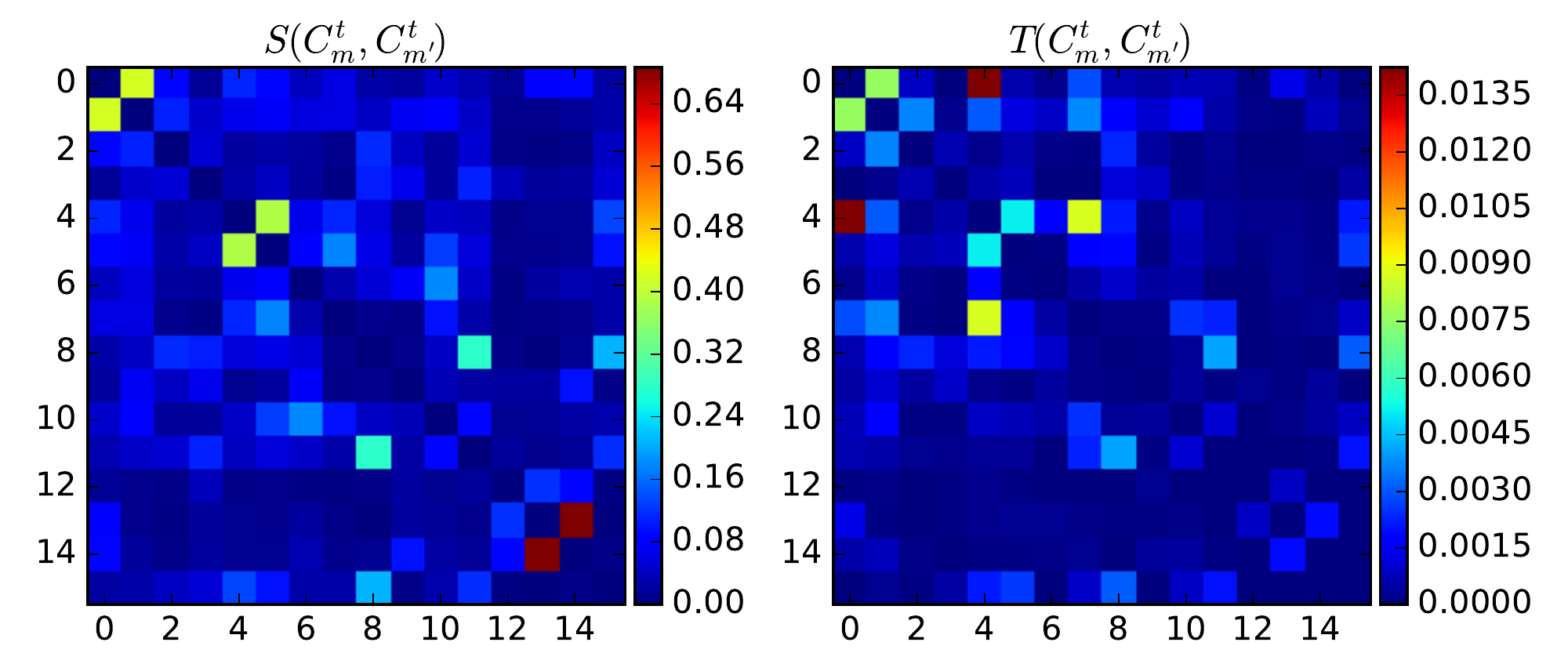}
  \caption{(left) $S(C_m^t, C_{m'}^t)$ of $16$ TCs in 1991, computed using forward intimacy indices going from 1991 to 1992. (right) $T(C_m^t, C_{m^\prime}^t)$ of the same $16$ TCs, using information from 1991 only. We use the same ordering of TCs in both matrices.}\label{fig:correlation}
\end{figure}

Next, we consider the Kuhnian processes of splitting and merging.
Unlike the ``prediction'' above, where we made use of information from years $t$ and year $t+1$, we would like to predict the splitting and merging of TCs using information only from year $t$.
Specifically, for merging events, our ground truth will be the similarity
\begin{equation}\label{eq:closeness}
  S(C_m^t, C_{m^\prime}^t) = \textstyle\sum_{n} (I_{mn}^f \mathbin{/} \textstyle\sum_{n'} I_{mn'}^f) (I_{m^\prime n}^f \mathbin{/} \textstyle\sum_{n''} I_{m^\prime n''}^f)
\end{equation}
between two TCs in year $t$, taking on values between $0$ and $1$.
If $C_m^t$ and $C_{m'}^t$ merge perfectly into a single TC in year $t+1$, $S=1$.
On the other hand, if the offsprings of $C_m^t$ and $C_{m'}^t$ are distinct, $S=0$.
To do prediction using only information from year $t$, we define
\begin{equation}\label{eq:linkage}
  T(C_m^t, C_{m^\prime}^t) = W(C_m^t, C_{m^\prime}^t) / (L(C_m^t)L(C_{m^\prime}^t)),
\end{equation}
where $W(C_m^t, C_{m^\prime}^t)$ is the sum of weights of edges between papers in $C_m^t$ and $C_{m'}^t$, normalized against the sizes of TCs involved.
\autoref{fig:correlation} shows that the two quantities are highly correlated, with Spearman's rank correlation coefficient of about $0.8$ (see \autoref{sec:merging}.)
A high $T$ between two TCs means they are likely to merge the next year.

We also tried to predict the splitting events.
Here the situation is more complex:
when we examine the weight matrix of a TC, we may find a few large subcommunities or many small subcommunites.
Naively, we expect the criterion for splitting is the opposite to merging, i.e. the easier it is to tell one subcommunity from others, the higher the chances for a split.
The \emph{boundary index}
\begin{equation}\label{eq:bi}
  B = \frac{\sum_{i_1 \neq i_2}\sum_{\substack{j_1 \in C_{i_1} \\ j_2 \in C_{i_2}}} A(j_1, j_2) / \sum_{i_i \neq i_2}L(C_{i_1})L(C_{i_2})}{\sum_{i} \sum_{j_1, j_2 \in C_i} A(j_1, j_2) / \sum_{i} L(C_i) L(C_i)}
\end{equation}
measures how indistinct the subcommunities are in a TC.
Here $A(j_1, j_2)$ is the weight of the edge between papers $j_1$ and $j_2$, and $C_i$ is a subcommunity in the given TC.
However the picture we find is not as simple as the merging case.
When we plot $M^f$ against $B$, we find the expected decreasing trend, but at the same time, the large scatter makes it impossible to reliably predict a splitting event using $B$.
To better understand the relationship between $M^f$ and $B$, we use quantile regression\cite{Sienkiewicz2016} to find that the $B$ has no ``prediction power'' when $M^f$ is small, but becomes ``predictive'' when $M^f$ is large.
We summarize these findings in \autoref{sec:splitting}
The slopes show that for the decile of most strongly splitting TCs, increasing the standardized $B$ by one standard deviation will decrease $M^f$ by about $0.05$, whereas for the decile of the least strongly splitting TCs, there is no obvious trend.
In \autoref{sec:splitting} we tested another index measuring a different network aspect of the weight matrix and found the prediction results are similar.

\begin{figure}
  \centering
  \includegraphics[width=\columnwidth]{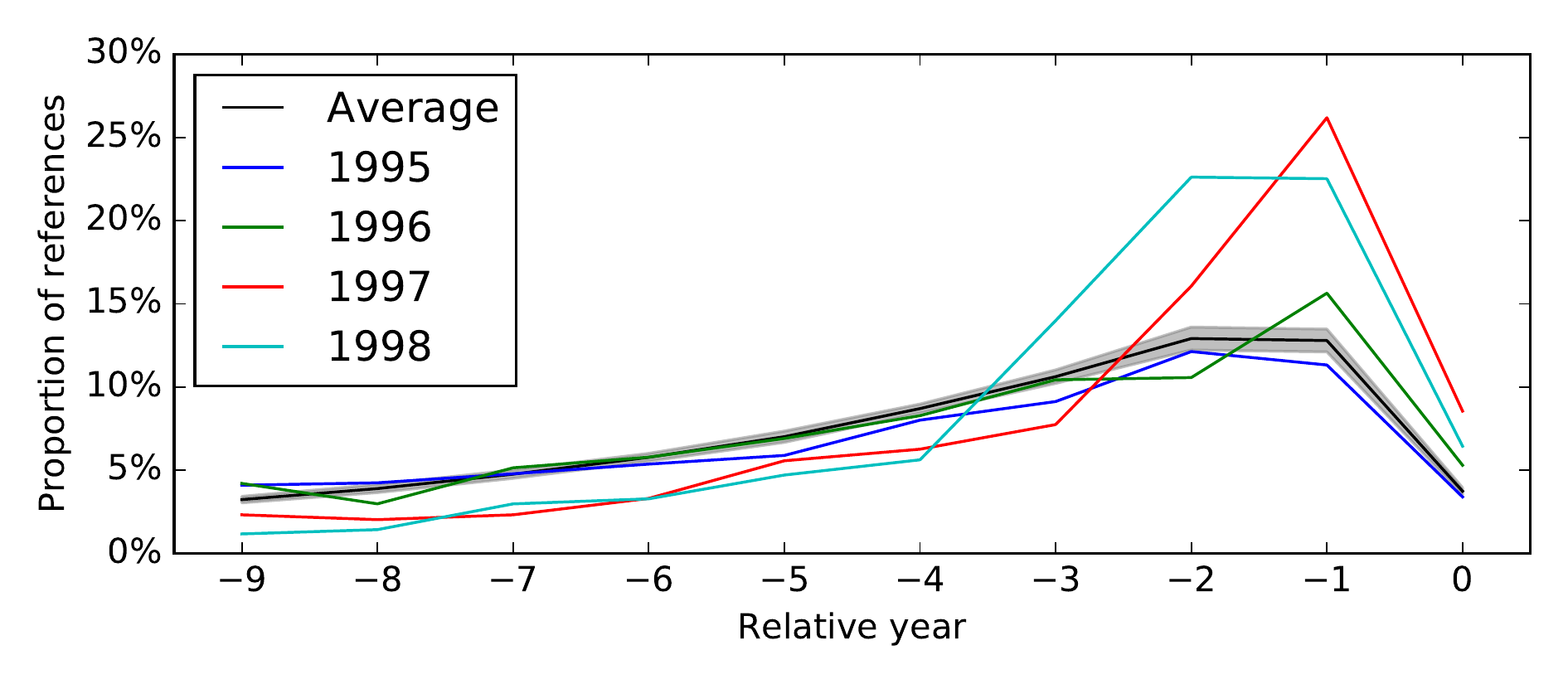}
  \caption{Proportions of a TC's references published in different years, relative to the year (0) of the TC. The black solid line is the proportions averaged over all TCs in the 1990s, while the area shaded gray is up to one standard deviation away from the mean. Other color lines represent the distribution of four different BEC related TCs.}\label{fig:sp}
\end{figure}

Finally, we want to know the impacts of such merging and splitting events.
We first check for an increase in the number of publications after such events, but found an insignificant difference in paper numbers in strongly and weakly mixing TCs (see \autoref{fig:sizerelation}(b) and (c)).
We suspected this is because our data set in confined to APS publications, and a more careful check should include other physics journals to capture any ``influence spillover".
We then check the highest, third quartile, median, average number of citations range two and five years after the events, but still see no significant effects (\autoref{sec:citation}).
Focusing on the highly productive chain of knowledge processes that led to experimental realizations of BEC, quantum teleportation and slow light, we checked the citation profiles between 1995 and 1998.
While the 1995 BEC+QI+QO TC cited a slightly lower proportion of 1995 papers than the APS 0-year average,
the 1996 BEC+QI+QO, the 1997 BEC TC, the 1998 BEC TCs all cited significantly more 0-year papers.
The full effect of this BEC breakthrough can be seen in the large proportions of 1996 papers cited by the 1997 and 1998 TCs and the proportion of 1997 papers cited by the 1998 TC (see \autoref{fig:sp}).
Indeed, we have provided early evidence suggesting that strongly mixing Kuhnian processes are associated with greater impact.

We thank Woo-Sung Jung for discussions.

%


\pagebreak
\widetext
\clearpage
\begin{center}
\textbf{\large Knowledge Evolution in Physics Research: An Analysis of Bibliography Coupling Networks \\
       Supplementary Material}
\end{center}
\setcounter{equation}{0}
\setcounter{figure}{0}
\setcounter{table}{0}
\setcounter{page}{1}
\makeatletter
\renewcommand{\theequation}{S\arabic{equation}}
\renewcommand{\thefigure}{S\arabic{figure}}
\renewcommand{\thetable}{S\arabic{table}}
\renewcommand{\bibnumfmt}[1]{[S#1]}
\renewcommand{\citenumfont}[1]{S#1}

\section{Null model of bibliography coupling network}\label{sec:nullmodel}
\begin{figure}[!h]
  \centering
  \includegraphics[width=0.8\textwidth]{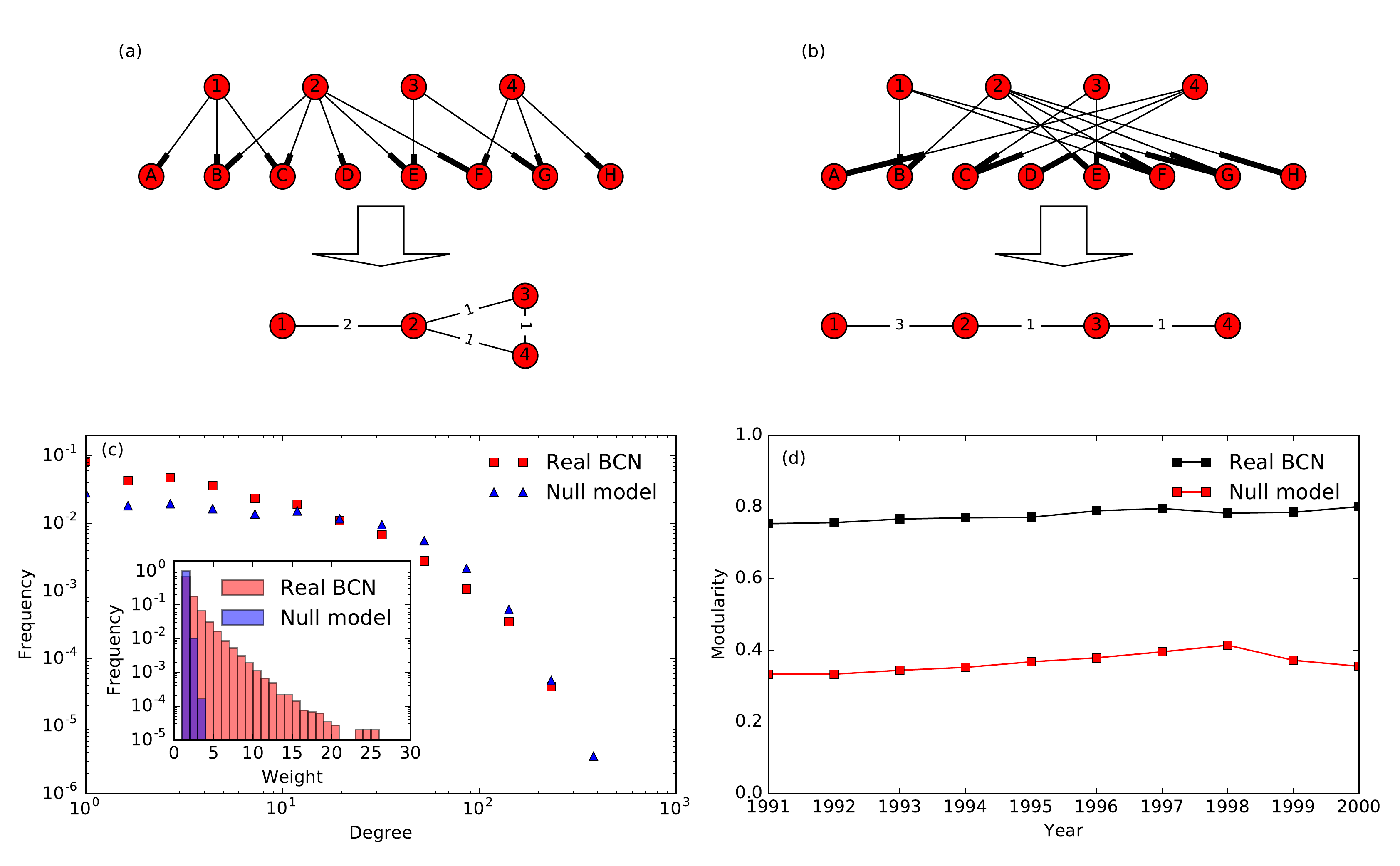}
  \caption{(a) Original citation network and its BCN. (b) A rewired citation network keeping in degrees and out degrees fixed and its BCN. (c) Comparison of the degree and weight distributions of papers published in 1991, between the real BCN and the null model. (d) Modularities of the best partitions extracted by the Louvain method for the real BCNs and the null model between 1991 and 2000. Results from null model are averaged over 10 different rewirings, and the error bars are much smaller than the marker size.}\label{fig:null_model}
\end{figure}

To determine the statistical significance of our empirical bibliographic coupling networks (BCNs) (\autoref{fig:null_model}(a)), we build a null model for comparison.
In our null model, we fix the out degrees and in degrees of all papers (citing and cited), but rewire the edges to get an ensemble of artificial BCNs (\autoref{fig:null_model}(b)).

Compared to the null model, the real BCN has more high-weight edges.
We suspect these are the most meaningful edges, arising from the paper's content.
If two papers focus on close topics, they will likely have high chance to have more than one common reference, and this effect also manifest itself in the degree distribution: the null model has a flatter degree distribution at small degrees because the edges are drawn by chance, whereas in the real BCN this coupling is based on content, meaning that papers will have edges mostly with papers that are trying to solve the same problems, so the real BCN will have more low-degree nodes, fewer high-degree nodes compared the null model.
The most prominent feature of this content-sensitive citation is community structure: in the real BCN, papers focussed on the same topic share more common references with each other than papers focussed on different topics, so that the densities of edges within topics are much higher than between topics. Therefore the modularities of communities extracted by the Louvain method in the real network is much higher than in the null model, as shown in \autoref{fig:null_model}(d).

\section{Validation of BCN communities}\label{sec:pacstest}
\begin{figure}[!h]
  \centering
  \includegraphics[width=0.8\textwidth]{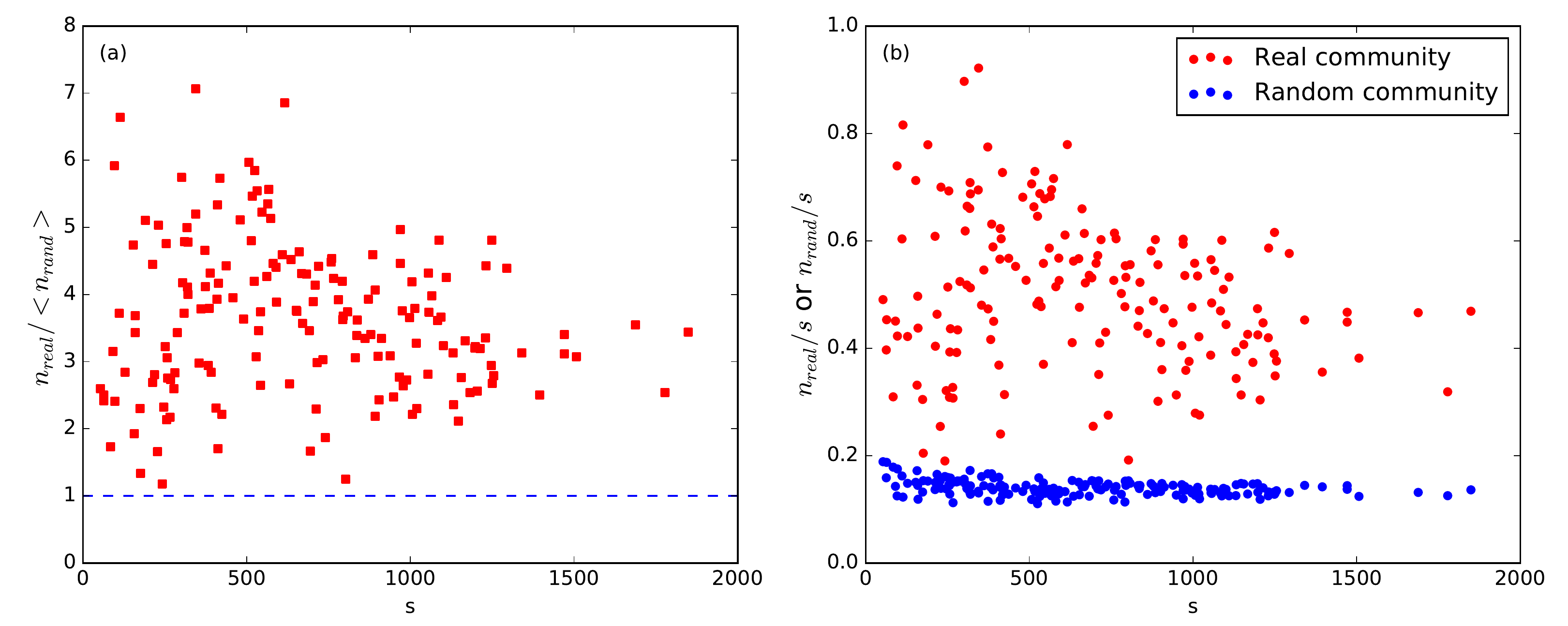}
  \caption{Comparison of PACS homogeneity between real BCN communities between 1991 and 2000 with more than 50 papers, and their corresponding random collections. (a) The red squares correspond to the sizes of the largest subsets of papers sharing at least one PACS number, $n_{real}$, in the empirical communities divided by the same quantity found in the corresponding random collections, $n_{rand}$, as a function of the community size $s$. (b) The fraction of the largest subset of papers sharing at least one PACS number as a function of $s$ for real communities in the BCN and random collections. For clarity, the small error bars are not shown in the figures.}\label{fig:homogeneity}
\end{figure}

To verify that the communities extracted are really focussed on closely related questions, we check the Physics and Astronomy Classification Scheme (PACS) numbers of members of the communities.
Such numbers are provided by authors to indicate the subfields of physics to which their papers belong.
In our case, we only use the first two digits of the PACS numbers, as a balance between accuracy and coverage.
To test whether the PACS numbers appearing in the communities could have occurred by chance, we choose one year $t$, build its BCN, extracting the community structure with sizes $\{s_1, s_2,..., s_n\}$, and then randomly assign papers in year $t$ into $n$ pseudo-communities of the same sizes, to remove any potential size effects.
For a community of size $s$, we then identify the largest subset of papers sharing the same PACS number.
This PACS number can represent the subfield of the community to a certain extent, and the fraction of papers in the largest subset reflect the homogeneity of the community.
On the other hand, the largest subset of papers sharing the same PACS number in a random collection of $s$ papers is typically small.
Dividing the sizes of the largest subsets in the empirical communities and in the random collections, we find ratios are larger than 1 for most cases.
This means that the communities we extracted are meaningful (see \autoref{fig:homogeneity}).

\section{Alluvial diagram for 1991-2000}\label{sec:prebirth}
\begin{figure}[!h]
  \centering
  \includegraphics[width=\textwidth]{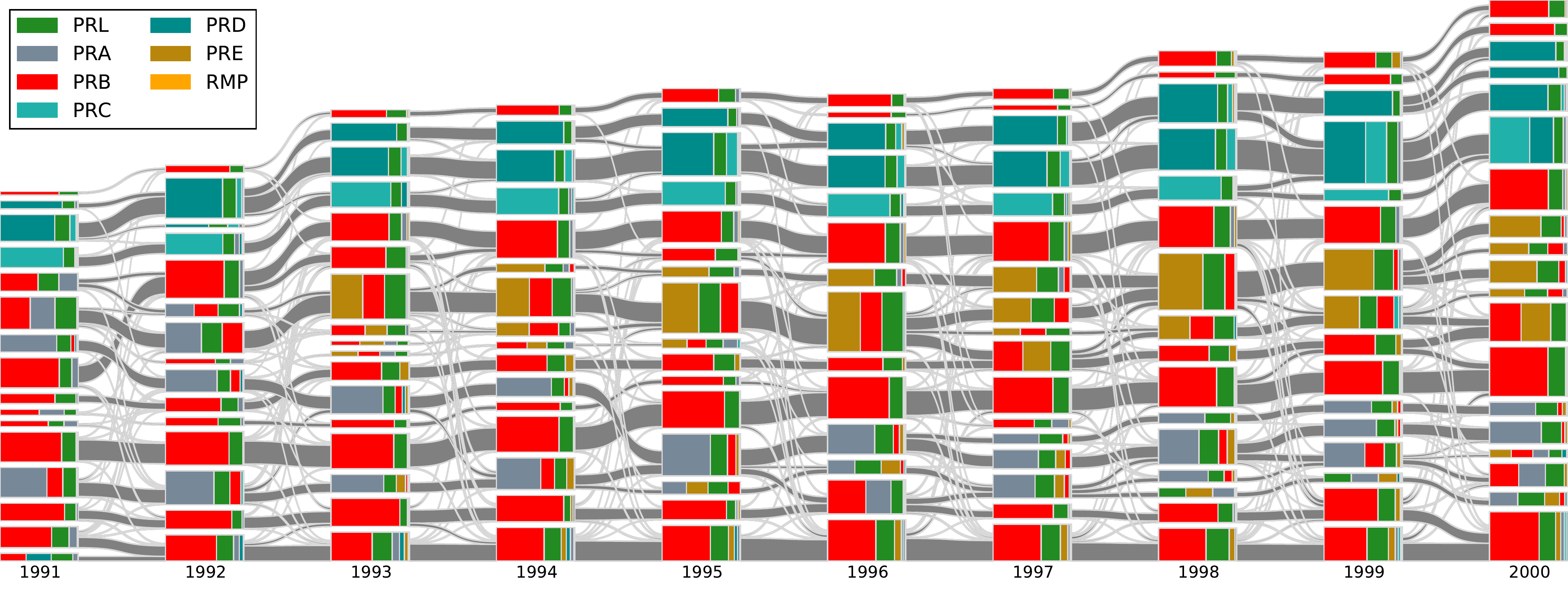}
  \caption{The alluvial diagram of APS papers from 1991 to 2000. Each block in a column represents a TC and the height of the block is proportional to the number of papers in the TC. Only communities comprising more than 100 papers are shown. TCs in successive years are connected by streams whose widths at the left and right ends are proportional to the forward and backward intimacy indices. The different colors in a TC represent the relative contributions from different journals.}\label{fig:alluvial diagram2}
\end{figure}

In addition to Fig. 2 in the letter, we also plotted an alluvial diagram for 1991 to 2000, showing the splitting of PRA into PRA and PRE.
As we can see from \autoref{fig:alluvial diagram2}, before 1993, there were several PRA-dominated TCs. After the split in 1993, some PRA-dominated TCs remained PRA-dominated, whereas other PRA-dominated TCs became PRE-dominated.
This means that even before 1993, papers in PRA were already divided into groups based on different topics, some of which are predecessors of the PRE TCs.

\section{Case study: quantum optics, quantum information and Bose-Einstein condensation}\label{sec:BEC}
\begin{figure}[!h]
  \centering
  \includegraphics[width=\textwidth]{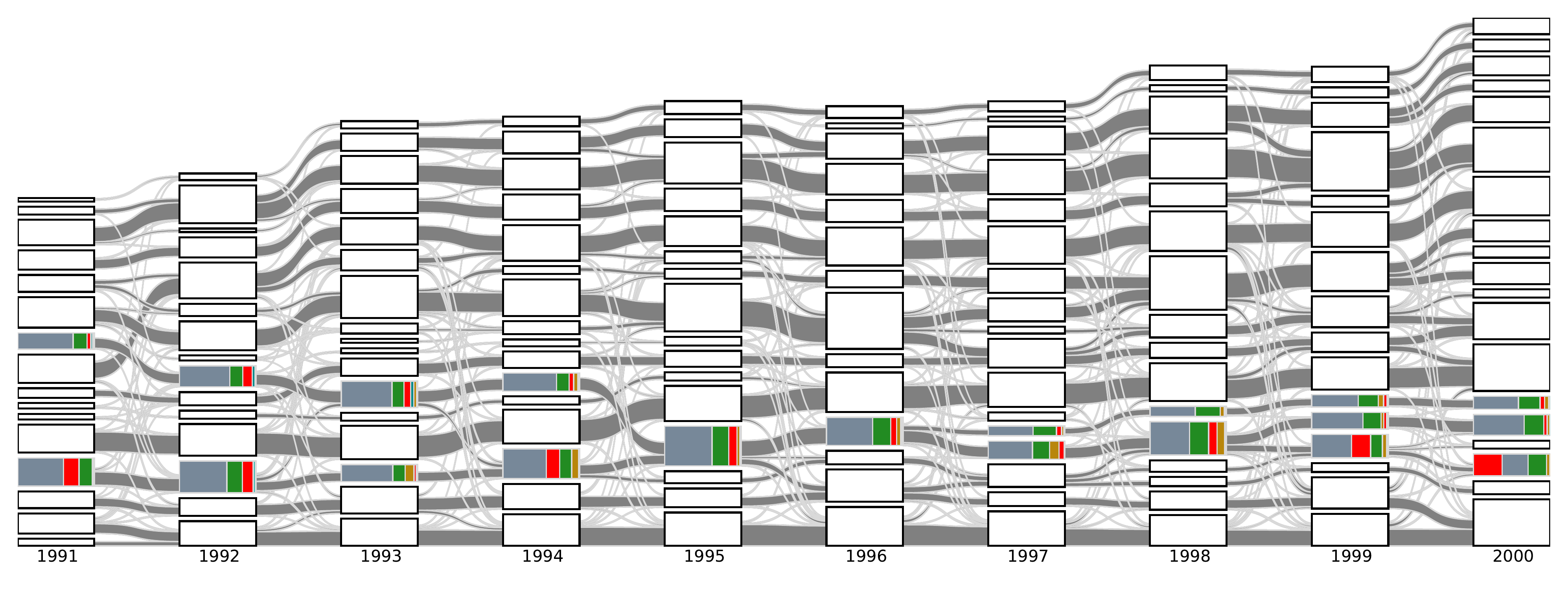}
  \caption{The alluvial diagram of APS papers from 1991 to 2000, where we colored only TCs highly related to quantum optics, quantum information and Bose-Einstein condensation.}\label{fig:BEC}
\end{figure}

To illustrate the utility our knowledge evolution framework can offer, we use as a case study the interesting interactions between quantum optics (QO), quantum information (QI), and Bose-Einstein Condensation (BEC).
These three fields experienced breakthroughs in the 1990s.
In \autoref{fig:BEC} we highlight the evolution of TCs which are related to these three topics and \autoref{tab:BEC} shows the three most cited papers in these TCs.
Key merging and splitting events are reported in the main paper, as are important publications these events are correlated with.

\begin{table}[!h]
  \centering
  \caption{The three most cited papers in quantum optics, quantum information theory, quantum computation and Bose-Einstein condensation related TCs. }\label{tab:BEC}
  \resizebox*{!}{\textheight}{%
  \begin{tabular}{|c|c|c|l|}
    \hline
    Year & TC & DOI & Title \\
    \hline
    \multirow{6}*{1991} & \multirow{3}*{Upper} & 10.1103/PhysRevLett.67.661 & Quantum cryptography based on Bells theorem \\
     &  & 10.1103/PhysRevLett.66.2593 & Observation of electromagnetically induced transparency \\
     &  & 10.1103/PhysRevLett.67.1855 & Enhancement of the index of refraction via quantum coherence \\
     \cline{2-4}
     & \multirow{3}*{Lower} & 10.1103/PhysRevA.44.5674 & \tabincell{l}{Above-surface neutralization of highly charged ions: The classical over-the-barrier \\ model} \\
     &  & 10.1103/PhysRevB.43.13401 & Strong magnetic x-ray dichroism in 2p absorption spectra of 3d transition-metal ions \\
     &  & 10.1103/PhysRevLett.66.2601 & Dynamic stabilization of hydrogen in an intense, high-frequency, pulsed laser field \\
    \hline
    \multirow{6}*{1992} & \multirow{3}*{Upper} & 10.1103/PhysRevLett.69.2881 & Communication via one- and two-particle operators on Einstein-Podolsky-Rosen states \\
     &  & 10.1103/PhysRevLett.69.3314 & \tabincell{l}{Observation of the coupled exciton-photon mode splitting in a semiconductor quantum\\ microcavity} \\
     &  & 10.1103/PhysRevLett.68.580 & Wave-function approach to dissipative processes in quantum optics \\
     \cline{2-4}
     & \multirow{3}*{Lower} & 10.1103/PhysRevLett.68.1943 & X-ray circular dichroism as a probe of orbital magnetization \\
     &  & 10.1103/PhysRevLett.68.3535 & High-order harmonic generation from atoms and ions in the high intensity regime \\
     &  & 10.1103/PhysRevLett.69.1383 & Absorption of ultra-intense laser pulses \\
    \hline
    \multirow{6}*{1993} & \multirow{3}*{Upper} & 10.1103/PhysRevLett.70.1895 & \tabincell{l}{Teleporting an unknown quantum state via dual classical and Einstein-Podolsky-Rosen\\ channels} \\
     &  & 10.1103/PhysRevA.47.4114 & Threshold and resonance phenomena in ultracold ground-state collisions \\
     &  & 10.1103/PhysRevLett.70.1244 & \tabincell{l}{Measurement of the Wigner distribution and the density matrix of a light mode using\\ optical homodyne tomography: Application to squeezed states and the vacuum} \\
     \cline{2-4}
     & \multirow{3}*{Lower} & 10.1103/PhysRevLett.71.1994 & Plasma perspective on strong field multiphoton ionization \\
     &  & 10.1103/PhysRevLett.70.1599 & Above threshold ionization beyond the high harmonic cutoff \\
     &  & 10.1103/PhysRevLett.70.774 & High-order harmonic generation in rare gases with a 1-ps 1053-nm laser \\
    \hline
    \multirow{6}*{1994} & \multirow{3}*{Upper} & 10.1103/PhysRevA.50.67 & Squeezed atomic states and projection noise in spectroscopy \\
     &  & 10.1103/PhysRevLett.72.3439 & Statistical distance and the geometry of quantum states \\
     &  & 10.1103/PhysRevLett.73.58 & Experimental realization of any discrete unitary operator \\
     \cline{2-4}
     & \multirow{3}*{Lower} & 10.1103/PhysRevA.49.2117 & Theory of high-harmonic generation by low-frequency laser fields \\
     &  & 10.1103/PhysRevLett.73.1227 & Precision Measurement of Strong Field Double Ionization of Helium \\
     &  & 10.1103/PhysRevA.50.1540 & Modeling harmonic generation by a zero-range potential \\
    \hline
    \multirow{3}*{1995} &  & 10.1103/PhysRevLett.75.3969 & Bose-Einstein Condensation in a Gas of Sodium Atoms \\
     &  & 10.1103/PhysRevLett.74.4091 & Quantum Computations with Cold Trapped Ions \\
     &  & 10.1103/PhysRevA.52.R2493 & Scheme for reducing decoherence in quantum computer memory \\
    \hline
    \multirow{3}*{1996} &  & 10.1103/PhysRevA.54.3824 & Mixed-state entanglement and quantum error correction \\
     &  & 10.1103/PhysRevLett.77.1413 & Separability Criterion for Density Matrices \\
     &  & 10.1103/PhysRevLett.77.2360 & Collective Excitations of a Trapped Bose-Condensed Gas \\
    \hline
    \multirow{6}*{1997} & \multirow{3}*{Upper} & 10.1103/PhysRevLett.78.985 & Bose-Einstein Condensation of Lithium: Observation of Limited Condensate Number \\
     &  & 10.1103/PhysRevLett.78.586 & Production of Two Overlapping Bose-Einstein Condensates by Sympathetic Cooling \\
     &  & 10.1103/PhysRevLett.78.5 & Demonstration of the Casimir Force in the 0.6 to 6$mu m$ Range \\
     \cline{2-4}
     & \multirow{3}*{Lower} & 10.1103/PhysRevLett.78.5022 & Entanglement of a Pair of Quantum Bits \\
     &  & 10.1103/PhysRevLett.78.3221 & \tabincell{l}{Quantum State Transfer and Entanglement Distribution among Distant Nodes in a \\ Quantum Network} \\
     &  & 10.1103/PhysRevLett.79.3306 & Noiseless Quantum Codes \\
    \hline
    \multirow{6}*{1998} & \multirow{3}*{Upper} & 10.1103/PhysRevLett.81.3108 & Cold Bosonic Atoms in Optical Lattices \\
     &  & 10.1103/PhysRevLett.81.938 & \tabincell{l}{Atomic Scattering in the Presence of an External Confinement and a Gas of \\ Impenetrable Bosons} \\
     &  & 10.1103/PhysRevLett.81.742 & Spinor Bose Condensates in Optical Traps \\
     \cline{2-4}
     & \multirow{3}*{Lower} & 10.1103/PhysRevA.57.120 & Quantum computation with quantum dots \\
     &  & 10.1103/PhysRevLett.80.2245 & Entanglement of Formation of an Arbitrary State of Two Qubits \\
     &  & 10.1103/PhysRevLett.81.5932 & \tabincell{l}{Quantum Repeaters: The Role of Imperfect Local Operations in Quantum \\ Communication} \\
    \hline
    \multirow{6}*{1999} & \multirow{3}*{Upper} & 10.1103/PhysRevLett.83.2498 & Vortices in a Bose-Einstein Condensate \\
     &  & 10.1103/PhysRevLett.83.5198 & Dark Solitons in Bose-Einstein Condensates \\
     &  & 10.1103/PhysRevLett.82.1975 & Entanglement of Atoms via Cold Controlled Collisions \\
     \cline{2-4}
     & \multirow{3}*{Middle} & 10.1103/PhysRevLett.83.4204 & Quantum Information Processing Using Quantum Dot Spins and Cavity QED \\
     &  & 10.1103/PhysRevB.59.2070 & Coupled quantum dots as quantum gates \\
     &  & 10.1103/PhysRevLett.82.2417 & Dynamical Decoupling of Open Quantum Systems \\
     \cline{2-4}
     & \multirow{3}*{Lower} & 10.1103/PhysRevLett.82.5229 & \tabincell{l}{Ultraslow Group Velocity and Enhanced Nonlinear Optical Effects in a Coherently \\ Driven Hot Atomic Gas} \\
     &  & 10.1103/PhysRevLett.83.2845 & Transmission Resonances on Metallic Gratings with Very Narrow Slits \\
     &  & 10.1103/PhysRevLett.83.967 & Liquid-Crystal Photonic-Band-Gap Materials: The Tunable Electromagnetic Vacuum \\
    \hline
    \multirow{9}*{2000} & \multirow{3}*{Upper} & 10.1103/PhysRevLett.84.806 & Vortex Formation in a Stirred Bose-Einstein Condensate \\
     &  & 10.1103/PhysRevLett.85.1795 & Stable 85Rb Bose-Einstein Condensates with Widely Tunable Interactions \\
     &  & 10.1103/PhysRevLett.85.3745 & Regimes of Quantum Degeneracy in Trapped 1D Gases \\
     \cline{2-4}
     & \multirow{3}*{Middle} & 10.1103/PhysRevA.62.062314 & Three qubits can be entangled in two inequivalent ways \\
     &  & 10.1103/PhysRevLett.84.2722 & Inseparability Criterion for Continuous Variable Systems \\
     &  & 10.1103/PhysRevA.62.012306 & \tabincell{l}{Electron-spin-resonance transistors for quantum computing in silicon-germanium \\ heterostructures} \\
     \cline{2-4}
     & \multirow{3}*{Lower} & 10.1103/PhysRevLett.85.5214 & Double Resonant Raman Scattering in Graphite \\
     &  & 10.1103/PhysRevLett.85.154 & Electronic Structure of Deformed Carbon Nanotubes \\
     &  & 10.1103/PhysRevB.62.13104 & Carbon nanotubes, buckyballs, ropes, and a universal graphitic potential \\
     \hline
  \end{tabular}%
  }
\end{table}

\section{Knowledge metabolism}\label{sec:metabolism}
\begin{figure}[!h]
  \centering
  \includegraphics[width=0.8\textwidth]{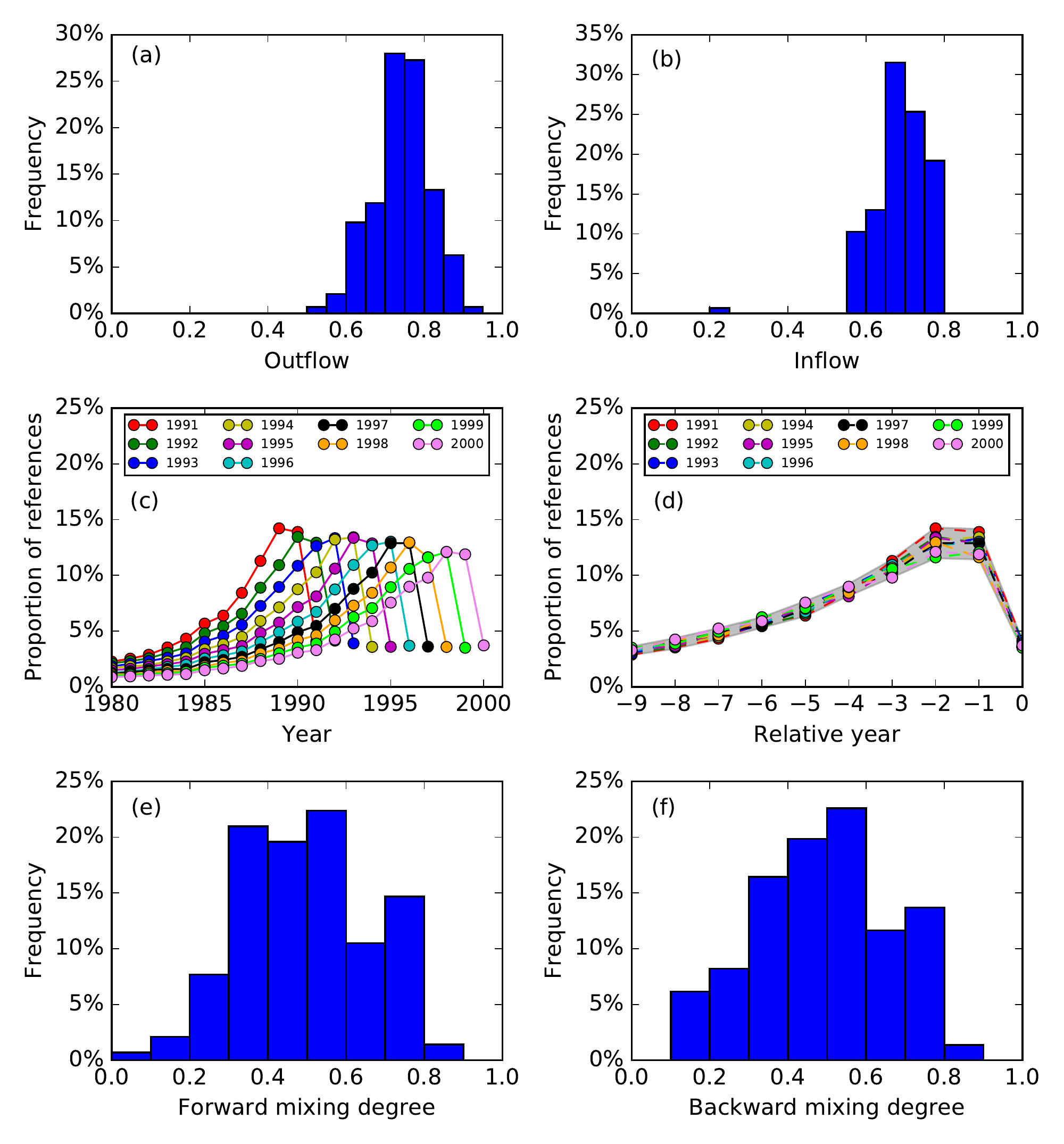}
  \caption{The metabolic analysis of APS papers in the 1990s. (a) The distribution of outflows of TCs. (b) The distribution of inflows of TCs. (c) Proportions of APS paper's references published in different years. (d) Proportions of APS paper's references published in different years, relative of the year (0) of publication. (e) The distribution of forward mixing degree of TCs. (d) The distribution of backward mixing degree of TCs.}\label{fig:metabolism}
\end{figure}

Some TCs have more references overlapping with those in the previous year, while other TCs have less.
To quantify evolution of references, we count the sums of the forward and backward intimacy indices.
These represent the percentage of references going to the next year, and the percentage of references inherited from the previous year, which we think of as the `outflow' and `inflow' respectively.
As shown in \autoref{fig:metabolism}(a) and (b), most outflows and inflows are distributed within a narrow range, but there are exceptional cases as well:
such as a single peak in \autoref{fig:metabolism}(b), whose references overlap significantly less than normal with the previous year.
In the context of birth, death, growth, decay, split, and merge knowledge processes, we are inclined to call this event in 1993 the birth of a TC.
Further analysis shows that most common PACS codes are: 03 (Quantum mechanics, field theories, and special relativity), 42 (Optics) and 63 (Lattice dynamics).
Looking at the references of this TC, we find that most of these comes from 1990, 3 year before.
This interesting phenomenon is therefore more appropriately identified as a sleeping beautify\cite{Ke2015}.

Every year, physicists absorb new references and drop old references as their fields progress.
Although this `metabolism' differ from TC to TC, the whole process is quite stable over all TCs, as shown in \autoref{fig:metabolism}(c) and (d).
This universal curve can be used as a benchmark for the test of scientific impact, as we have done in \autoref{fig:sp} of the letter.

As we see from \autoref{fig:alluvial diagram2}, there is a great diversity of processes acting on the TCs: some TCs are derived almost exclusively from one source, others receive strong contributions from a small number of sources, or weak contributions from a large number of sources. To quantify such diversity, we introduce forward mixing degree and backward mixing degree (\autoref{eq:mixingdegree} in the letter)
As shown in \autoref{fig:metabolism}(e), (f), strong splitting/merging or almost isolated development are rare, and in most case, TCs undergo weak information exchange.

\section{Prediction of merging}\label{sec:merging}
\begin{figure}
  \centering
  \includegraphics[width=0.5\textwidth]{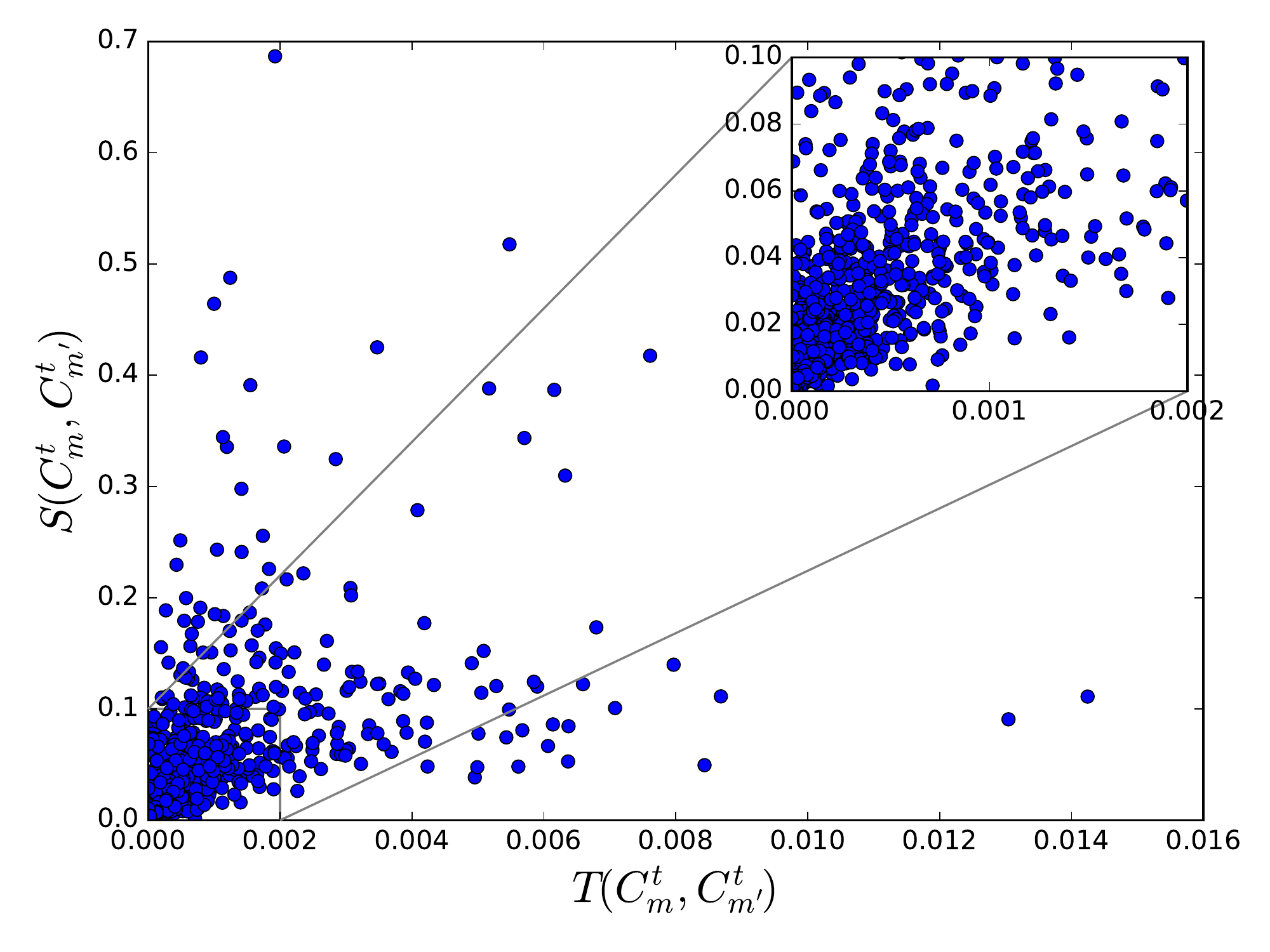}
  \caption{The scatter plot between $T(C_m^t, C_{m'}^t)$ and $S(C_m^t, C_{m'}^t)$ among all TCs (with at least 100 papers) in 1990s.}\label{fig:pm}
\end{figure}

As mentioned in the letter, we found that the inter-TC connection is highly correlated with their mixing the next year.
High $T(C_m^t, C_{m'}^t)$ leads with a large probability to a high $S(C_m^t, C_{m'}^t)$.
Analyzing APS papers in the 1990s, we found a Spearman's rank coefficient of $0.804$ between $T(C_m^t, C_{m'}^t)$ and $S(C_m^t, C_{m'}^t)$ over all TCs (with at least 100 papers).
However, because the average Pearson correlation coefficient is only $0.504$, such a relation is not linear (see \autoref{fig:pm})

\section{Hierarchy structure of community and splitting analysis}\label{sec:splitting}
\begin{figure}
  \centering
  \includegraphics[width=\textwidth]{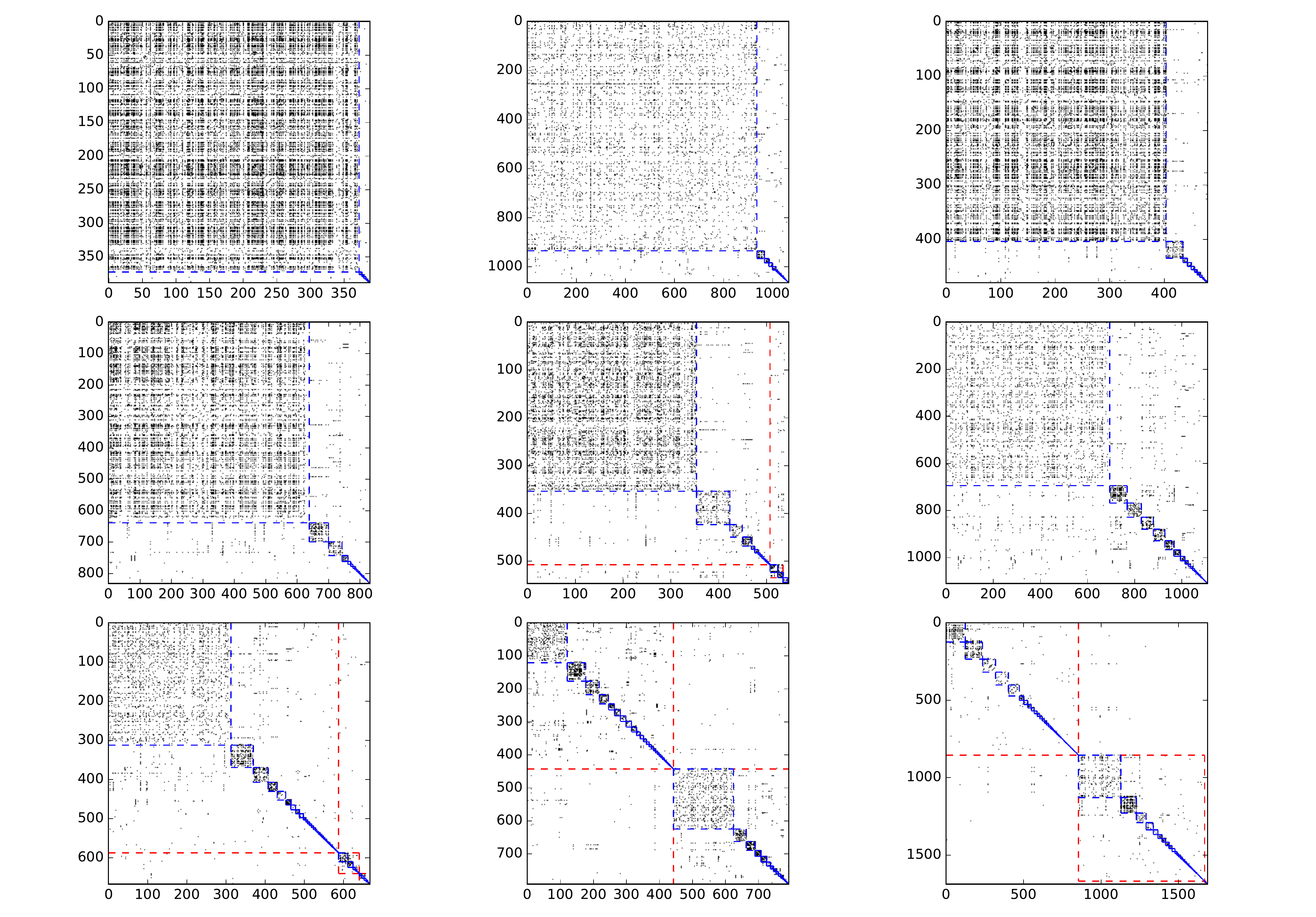}
  \caption{Adjacency matrices of TCs in the 1990s. The blue lines indicate the boundaries of subsubcommunities, the red lines indicate the boundaries of subcommunities. The red lines are absent from some plots because such TC have only one level when the Louvain algorithm terminated.}\label{fig:hierarchystructure}
\end{figure}

To predict a splitting event, we check the weight matrix of different TCs.
Naively, we expect the components that lead on to different TCs the next year to already form distinct subcommunities this year.
However, when we use the dendrogram extracted from the Louvain method to identify subcommunities, we found that different TCs have different internal structures (see \autoref{fig:hierarchystructure}): some have a few large subcommunities, while others have many small subcommunities.

\begin{figure}
  \centering
  \includegraphics[width=\textwidth]{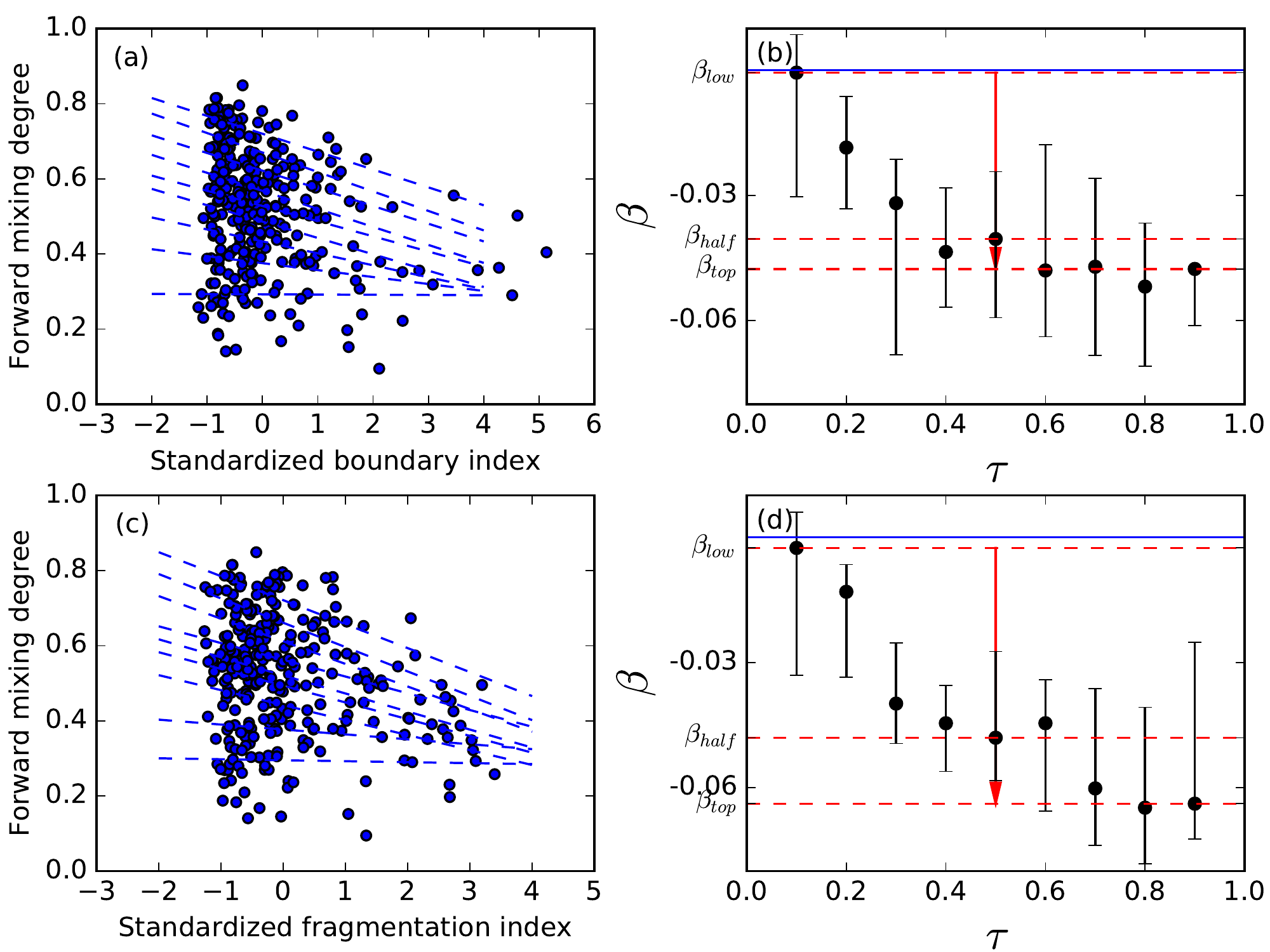}
  \caption{Relation between boundary index, fragmentation index and forward mixing degree of TCs in 1980s and 1990s. (a) Each dot corresponds to one TC, dash lines show QR results for quantiles $\tau = {0.1, 0.2, ..., 0.9}$. (b) $\beta$ coefficients (slopes of QR in the (a) as a function of $\tau$. The red arrows show $\beta_{low} \equiv \beta (\tau = 0.1)$, $\beta_{half} \equiv \beta (\tau = 0.5)$ and $\beta_{top} \equiv \beta (\tau = 0.9)$, as, respectively, the nock, a circle on the shaft, and the head of the arrow, the blue solid line represents $0$. (c) Each dot corresponds to one TC, dash lines show QR results for quantiles $\tau = {0.1, 0.2, ..., 0.9}$. (d) $\beta$ coefficients (slopes of QR in the (c) as a function of $\tau$. The red arrows show $\beta_{low} \equiv \beta (\tau = 0.1)$, $\beta_{half} \equiv \beta (\tau = 0.5)$ and $\beta_{top} \equiv \beta (\tau = 0.9)$, as, respectively, the nock, a circle on the shaft, and the head of the arrow, the blue solid line represents $0$.}\label{fig:qr}
\end{figure}

Assuming that splitting is the time reversal of merging, i.e. two distinct subcommunities becoming two distinct communities in the next year, we devise the boundary index $B$ to measure how indistinct the subcommunities are in a TC.
Quantile regression (QR) shows that the relation between $B$ and $M^f$ depends on the decile, as shown in \autoref{fig:qr}(a), (b).
For different decile, the relation is different.

We also try with \emph{fragmentation index}
\begin{equation}\label{eq:fi}
  F = \sum_{i:j[i]} w_i S_{j[i]}^2
\end{equation}
where $w_i$ is the size fraction of the top level subcommunity $i$, $s_{j[i]}$ is the relative size fraction of subsubcommunity $j$ inside subcommunity $i$.
The more fragmentation a community is, the closer $F$ is to $0$.
Quantile regression between $F$ and $M^f$ gives very similar results as $B$ and $M^f$, that is, for the decile of most strongly splitting TCs, increasing the standardized $F$ by one standard deviation will decrease $M^f$ by about 0.06, whereas for the decile of the least strongly splitting TCs, there is no obvious trend as $\beta$ close to $0$, as shown in \autoref{fig:qr}(c), (d).

\section{Correlation analysis between citation and mixing}\label{sec:citation}
\begin{figure}
  \centering
  \includegraphics[width=0.8\textwidth]{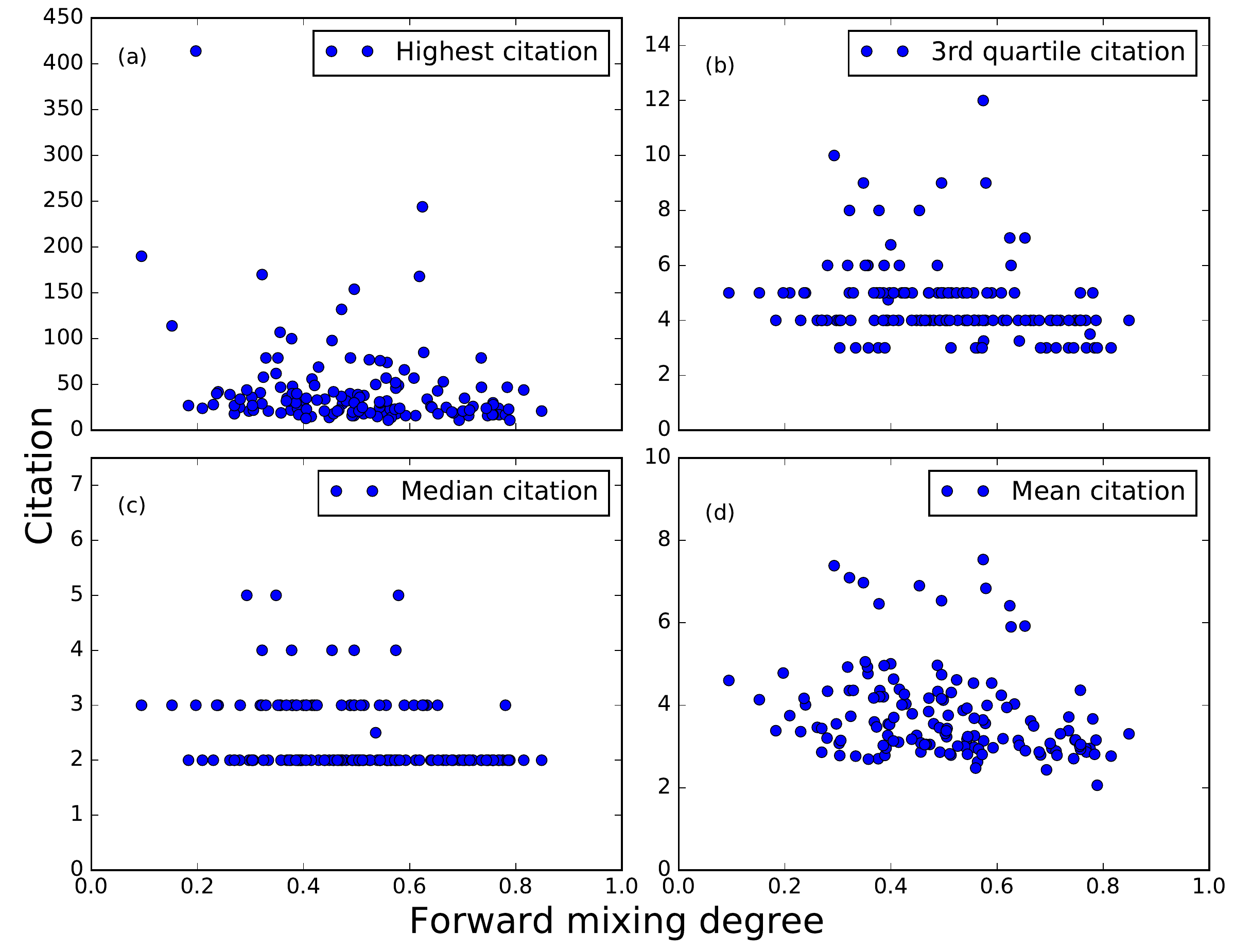}
  \caption{The scatter plot between different citations received during 2 years and forward mixing degree among all TCs (with at least 100 papers) in 1990s. (a) Highest citation, (b) Third quartile citation, (c) Median citation, (d) Mean citation.}\label{fig:citation}
\end{figure}

When we think of high-impact research, we think of highly-cited papers.
Therefore, to quantify the impact of strongly-splitting events in the alluvial diagrams, we counted the citations of TCs resulting from splittings.
As shown in \autoref{fig:citation}, we did this for number of citations 2 years after the events, and also 5 years after the events.
There were no obvious trends.
The results of backward mixing degree, i.e. merging, are similar.

\end{document}